\documentclass[
  aps,prb,
  showpacs,
  twocolumn,
  amsmath,
  amssymb,
  floatfix,
  superscriptaddress, 
  longbibliography
]{revtex4-1}

\usepackage{newtxtext,newtxmath}

\usepackage{hyperref}
\hypersetup{colorlinks=true,urlcolor=blue,citecolor=blue,linkcolor=blue,breaklinks=true}

\usepackage{
  graphicx,
  cleveref,
  sidecap,
  xargs
}

\setcitestyle{super}

\graphicspath{{Figures/}}
\newcommand{\reff}[1]{Fig.~\ref{fig:#1}}
\newcommand{\refq}[1]{Eq.~(\ref{eq:#1})}
\newcommand{\refs}[1]{Sec.~(\ref{sec:#1})}

\creflabelformat{section}{(#1#2#3)}
\creflabelformat{figure}{(#1#2#3)}
\Crefname{equation}{Eq.}{Eqs.}
\Crefname{section}{Sec.}{Secs.}
\Crefname{figure}{Fig.}{Figs.}
\Crefname{table}{Tab.}{Tabs.}

\newcommand{\Chi}{\protect\raisebox{2pt}{$\chi$}}
\newcommand{\vek}[1]{\mathbf{#1}}
\newcommand{\vesan}[3]{{#1}^{#2}_{#3}}

\newcommand{\crea}[1]{c^{\dagger}_{#1}}
\newcommand{\anni}[1]{c^{\phantom{\dagger}}_{#1}}
\newcommand{\up}{\uparrow}
\newcommand{\dn}{\downarrow}
\newcommand{\ver}{\text{ver}}
\newcommand{\bub}{\text{bub}}

\newcommand{\pade}{Pad\'{e}}
\newcommand{\victory}{{\em victory }}

\newcommand{\twop}{two-particle}
\newcommand{\imag}{\text{Im}}

\newcommand{\gsim}{\gtrsim}
\newcommand{\lsim}{\lesssim}

\makeatletter
  \def\l@subsubsection#1#2{}%
\makeatother

\begin{document}

  \title{Competition between antiferromagnetic and charge density wave fluctuations in the extended Hubbard model}
 
  \author{Petra Pudleiner}
  \affiliation{Institute for Solid State Physics, Vienna University of Technology, 1040 Vienna, Austria}
  \author{Anna Kauch}
  \affiliation{Institute for Solid State Physics, Vienna University of Technology, 1040 Vienna, Austria}
  \author{Karsten Held}
    \email{held@ifp.tuwien.ac.at}
  \affiliation{Institute for Solid State Physics, Vienna University of Technology, 1040 Vienna, Austria}
  \author{Gang Li}
  \email{ligang@shanghaitech.edu.cn}
 \affiliation{School of Physical Science and Technology, ShanghaiTech University, Shanghai 200031, China}
 \affiliation{\mbox{ShanghaiTech Laboratory for Topological Physics, ShanghaiTech University, Shanghai 200031, China}}

  \date{\today}
  
  \begin{abstract}
By extending our {\it victory} implementation of the parquet approach to include  non-local Coulomb interactions, we study the extended Hubbard model on the two-dimensional square lattice with a particular focus on the competition of the non-local charge and spin fluctuations.
Surprisingly, we find that their competition, as the mechanism driving the phase transition towards the charge density wave, dominates only in a very narrow parameter regime in the immediate vicinity of the phase transition. 
Due to the special geometry and the Fermi surface topology of the square lattice,  antiferromagnetic fluctuations  dominate even for sizable next-nearest neighbor interactions. 
Our conclusions are based on the consistent observations in both the single- and two-particle quantities, including the self-energy, the single-particle spectral function, the two-particle susceptibility, the density-density vertex function and the optical conductivity. 
Our work unbiasedly establishes the connection of these quantities to the charge fluctuations, and the way of interpretation can be readily applied to any many-body method with access to the two-particle vertex. 
  \end{abstract}
  \maketitle

\section{Introduction}
\label{sec:Introduction}
The study of strongly correlated electron systems is notoriously difficult and represents one of the greatest challenges in contemporary condensed matter physics. 
The competition of the various entangled degrees of freedom, such as the charge, spin and orbital contributions, is often the source of emerging exotic physics. On the other hand, it imposes a great challenge to theory.
The interest in  correlated electron models, especially the Hubbard model with an interaction strengths comparable to the bandwidth, is highly motivated by the unconventional physics beyond the simple itinerant or localized picture discovered in high-temperature superconductors \cite{Bednorz1986, Schiilling, ANDERSON1196, RevModPhys.70.1039, Dagotto94, Orenstein468}. 
The half-filled, two-dimension Cu-O plane shows magnetic long-range order, which can be gradually removed by electron- or hole-doping,  necessary for the emergence of superconductivity. 

Besides magnetic order and fluctuations also the relevance of charge density wave fluctuations and order is  discussed---already in the early days of high-temperature superconductivity~\cite{Bickers87} but more intensively after experimental evidence thereof \cite{Wu2011,Chang2012,Ghiringhelli2012}. These charge-density wave fluctuations are fostered by non-local Coulomb interactions,
 leading to the so-called extended Hubbard model (EHM)~\cite{Devreese, Solyom, PhysRevLett.93.086402,PhysRevLett.77.3391,PhysRevB.66.085120,PhysRevLett.92.196402} as a minimal model. The EHM Hamiltonian reads
\begin{align}
H=
   &-t\sum_{\langle ij \rangle,\sigma} \crea{i\sigma}\anni{j\sigma}- \mu\sum_{i,\sigma}n_{i\sigma} \nonumber \\
   &+ U \sum_{i} n_{i\up}n_{i\dn} + \frac{V}{2} \sum_{\langle ij \rangle, \sigma\sigma'} n_{i\sigma}n_{j\sigma'}\;,
   \label{eq:Hamilton}
\end{align}
where $c_{i\sigma}^{(\dagger)}$ is the annihilation (creation) of electrons with spin $\sigma\in\{\up,\dn\}$ at lattice site $i$.
The particle density is hence given by $n_{i\sigma}=\crea{i\sigma}\anni{i\sigma}$.
The kinetic term of the Hamiltonian \eqref{eq:Hamilton} describes the hopping of electrons with amplitude $t$ to neighboring sites only, via $\langle ij \rangle$; 
The screened long-range Coulomb potential is described by an on-site density-density interaction with strength $U$ and by a $V$-term including interaction between electrons on neighboring sites.
 In this paper, we restrict ourselves to  the case of only positive $U$ and $V$ since the Coulomb interaction is repulsive, albeit low-energy effective models may also be attractive and show interesting physics~\cite{RevModPhys.62.113}.

As already mentioned, the presence of non-local interaction term introduces, in addition to the antiferromagnetic correlation caused by the on-site interaction, the intersite charge-density correlations.
The competition of the two types of correlations leads to a phase transition between the spin-density wave (SDW) and the charge-density wave (CDW) in the low-temperature regime.  
From the low-energy effective theory point of view, a weak non-local interaction can be viewed as an additional screening of the local interaction. 
An effective local interaction, after integrating out the non-local interaction, can then be obtained.  
However, the intersite fluctuations (or the non-local charge fluctuations) are absent in such effective theory.
That is, in the effective theory the bare interaction is renormalized by the frequency/energy dependent dielectric function determined from the charge-charge correlation function and the effective local interaction becomes dynamic, {\it i.e.} $U(\omega)$~\cite{PhysRevB.70.195104, PhysRevB.71.035105, PhysRevB.87.165118, PhysRevB.85.045132}
With the increase of the non-local interactions, the ''screening" picture is, however, no longer appropriate due to the underestimation of the non-local interaction effect and a more complete description of the two interactions in the EHM is required. 

It is, however, difficult to treat the two types of interaction in Eq.~(\ref{eq:Hamilton}) simultaneously. Let us start with the exact solution in the atomic limit, {\it i.e.} for $t_{ij}=0$. 
In one dimension at half-filing, the ground states of the atomic EHM is in a CDW ordered phase when $U<0$ irrespective the value of $V$.  
This phase is characterized by the alternative appearance of the empty and the doubly occupied states, forming the CDW with two-site periodicity. 
It extends to the positive value of $U$ as long as $U/2 < V$, otherwise the Mott phase is realized where only singly occupied states are allowed~\cite{PhysRevE.77.061120}. 
Similar analysis can be carried over to higher dimensions, and the transition to the  CDW phase was found to occur at $U=2n_{dim}V$, where $n_{dim}$ is the system dimensionality~\cite{DongenStrCoup,Bari1971}. 

\begin{figure}[htbp]
   \includegraphics[width=0.9\linewidth]{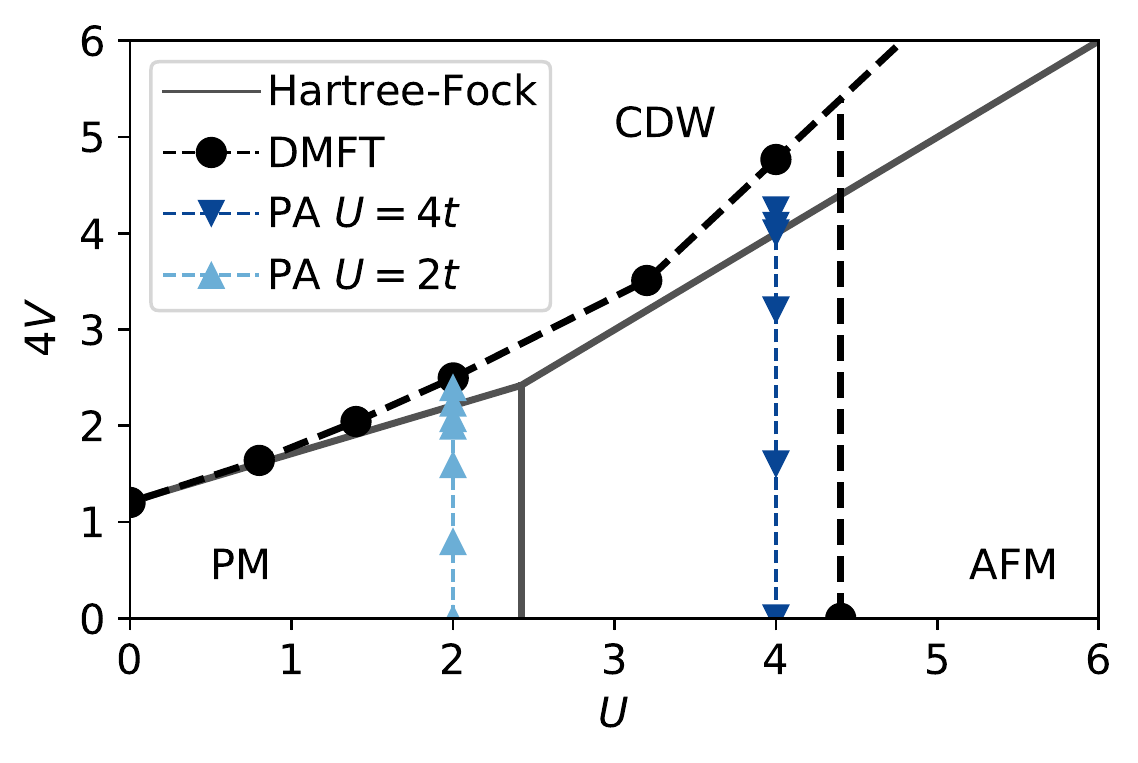}
   \caption{Phase diagram of the EHM in the Hartree-Fock,\cite{Wahle1998} (solid line) and the DMFT\cite{Wahle1998} (dashed line with circles) approximations. Antiferromagnetic, charge-density wave and paramagnetic solution are denoted as AFM, CDW and PM. The last occurs because of the finite temperature, {\em i.e.} $T=1/4t$ in the DMFT calculation.\cite{Wahle1998,note1} The triangles represent the parameters at which the {calculations} of the present paper were performed using the parquet approximation  (denoted as PA, light blue for $U=2t$ and dark blue for $U=4t$).
}
   \label{fig:PhaseWahle}
\end{figure}

Wherever the electronic hopping process is recovered, the EHM loses its exactly solvable limit. Either approximations or numerical methods have to be consulted~\cite{PhysRevB.39.9397, PhysRevB.49.7904, PhysRevB.50.14016,Kapcia2011,PhysRevB.96.205102}. 
Figure~\ref{fig:PhaseWahle} displays the phase diagram obtained in the Hartree-Fock approximation and the dynamical mean-field theory (DMFT) for the two-dimensional EHM~\cite{Wahle1998}.
A strong coupling expansion around the atomic limit predicts the transition to occur at values of $V>U/4$, \cite{DongenStrCoup} {\em i.e.} for $V$'s larger than in the atomic limit.
In DMFT, the non-local interaction $V$ reduces to its Hartree contribution \cite{Muller-Hartmann1988} but the difference   to the Hartree-Fock solution of the EHM is that the local interaction $U$ remains dynamic. 
This DMFT solution~\cite{Wahle1998} predicts the transition at a larger $V$ value than in the Hartree-Fock or atomic limit solution and it increases with the increase of $U$ as shown in Fig.~\ref{fig:PhaseWahle}. 
The light/dark blue triangles in \reff{PhaseWahle} denote the parameters at which our calculations were performed, which will be discussed later in Section~\ref{sec:Results}.
The shift of the phase transition line towards larger value of $V$ was confirmed by exact numerical simulations of DQMC~\cite{Zhang1989}.

In contrast to the DMFT, the extended DMFT (EDMFT)~\cite{edmft1} contains the non-local $V$-term in its construction and treats it self-consistently in the effective polarization function. 
In EDMFT the transition was also found at $4V/U>1$ in the strong coupling regime, which is similar to the DMFT solution.
The similarity is partially due to the fact that the EDMFT approximation also neglects the non-local fluctuations which become crucial in low-dimensional systems.
The diagrammatic extensions of EDMFT, {\it i.e.} the dual boson approach~\cite{Vanloon2014}, captures the non-local fluctuations and hints to a transition closer to $4V/U=1$ (still however with $4V/U>1$) in the intermediate coupling regime as well.
The transition boundary in the dual boson approach lies in between the solid and dashed lines in Fig.~\ref{fig:PhaseWahle}.

To go beyond the DMFT/EDMFT local approximations, either a cluster \cite{RevModPhys.77.1027}  or a diagrammatic extension \cite{RohringerRMP} of DMFT is required. 
In a recent dynamical cluster approach (DCA) study~\cite{Gull2017,Paki2019}, a similar tendency was confirmed. 
In this study, a momentum cluster of 20-sites was considered where the short-range fluctuations within the cluster were fully taken into account. Note, however, that for the larger values of $U$ used in this DCA study, the system without or with weak non-local interaction $V$ is in the ordered antiferromagnetic (AFM) phase, which is a finite-size effect. This AFM ordering is suppressed within the parquet approach used here already for the finite lattices considered.

The aim of this paper is to provide an unbiased study of the EHM with both local and short-range fluctuations included in a way that both the single- and two-particle Green's functions are self-consistently determined. We study, for a wide range of non-local interactions, the competition between magnetic and charge fluctuations before the phase transition. Both the one- and two-particle quantities indicate that the charge fluctuations dominate only in the vicinity of the transition, leading to a suppression of the AFM fluctuations. But otherwise, even for sizable non-local interactions, AFM fluctuations prevail. The precise transition boundary is not of the prime interest of this work, instead we want to show how the transition is approached and how it is probed in the parquet approach, which is also of  interest to many other cluster methods that have the access to the two-particle vertex functions. This work also represents a methodological progress which further extends the {\it victory}  solver of the parquet equations~\cite{Bickers2004, victory1, victory2}  to include the non-local Coulomb interactions in a lattice model.

Throughout the paper, $t\equiv1$, $\hbar\equiv1$, and $k_B\equiv1$ are set as the units of energy, frequencies and temperature.
The square lattice with average number of electrons per site $n\equiv1$ will be studied at temperature $T=t/6$ and onsite interactions are taken as $U=2t$ or $U=4t$. 
The nearest-neighbor interaction $V$, which reads $V_{\vek{q}}=2V[\cos(q_x)+\cos(q_y)]$ in momentum space,  will be varied as fractions of $U/4$ for a given $U$ to gradually approach the transition.

Section 
\ref{sec:ParquetMethod} recapitulates the parquet equations and explains the methodological development of the present paper. Section~\ref{sec:Results} presents the results obtained, except for the  optical conductivity which is discussed in Section~\ref{sec:OptCondu}.

\section{Method: parquet approximation to the EHM}
\label{sec:ParquetMethod}
First, we will explain the methodological development of the {\it victory} package~\cite{victory2} with respect to the non-local interaction. 
The extension is straightforward in the sense that the kernel approximation, which is the key concept of the {\it victory} implementation of the parquet approach, can still be applied. 
However, due to the non-local interaction the various two-particle vertex functions become more strongly momentum-dependent. 
Thus, one has to introduce the two-level kernel approximations~\cite{victory1} in each momentum patch, as will be explained in this section.

In the parquet approach, a set of exact equations  that couple the one- and \twop{} vertex functions is solved iteratively until a self-consistency in both levels is achieved. 
This is conceptually different from any single-particle theory, where  only the self-consistency imposed by the Dyson equation, such as the DMFT. 
In the parquet approach, all the single-particle quantities including the Green's function and the self-energy are determined by the two-particle vertex functions in all fluctuating channels. 
As a result,  it allows for an unbiased treatment of all fluctuations simultaneously.
This is also in sharp contrast to any ladder approximation where only a set of {\em a priori} chosen ladder diagrams are considered in favor of only certain type of fluctuations.    
More details about the parquet approach can be found in references~\cite{PhysRevB.46.8050, PhysRevB.47.8851, PhysRevB.75.165108, PhysRevB.83.035114, Janis2017, Chen1992311, PhysRevB.43.8044, Bickers2004, PhysRevE.80.046706, PhysRevE.87.013311, victory1, victory2}. 
Here we would like to concentrate on the extension of it to the EHM.
Thus, only the most relevant formulas will be given.   

In the parquet approach, the only input needed is the fully irreducible \twop{} vertex $\Lambda^{kk'q}$.
Here, the four-vector notation with $k=(\vek{k},\nu_n)$ and $q=(\vek{q},\omega_n)$ is used with the momenta
$\vek{k}$ and $\vek{q}$, the discrete Matsubara frequencies are 
$\nu_n=\frac{\pi}{\beta}(2n+1)$ (fermionic) and 
$\omega_n=\frac{\pi}{\beta}2n$ (bosonic) with $n\in\mathbb{Z}$ and the inverse temperature is $\beta=1/T$.
In this work we employ the parquet approximation (PA),\cite{Bickers2004} in which the fully irreducible \twop{} vertex $\Lambda^{kk'q}$ is approximated by its frequency independent lowest-order contribution, namely the bare interaction.
Compared to the Hubbard model, the inclusion of the nearest-neighboring Coulomb interaction $V$  leads to additional terms in the lowest-order vertex functions:
\begin{alignat}{1}
  \vesan{U}{\vek{k}\vek{k}'\vek{q}}{d}
  &= U + 2V_{\vek{q}} - V_{\vek{k}'-\vek{k}}\;,
  \label{eq:PAd}\\
  \vesan{U}{\vek{k}\vek{k}'\vek{q}}{m}
  &=-U - V_{\vek{k}'-\vek{k}}\;,
  \label{eq:PAm}\\
  \vesan{U}{\vek{k}\vek{k}'\vek{q}}{s}
  &=-2U - V_{\vek{k}'-\vek{k}} - V_{\vek{q}-\vek{k}-\vek{k}'},
  \label{eq:PAs}\\
  \vesan{U}{\vek{k}\vek{k}'\vek{q}}{t}
  &= V_{\vek{k}'-\vek{k}} - V_{\vek{q}-\vek{k}-\vek{k}'},
  \label{eq:PAt}
\end{alignat}
in the respective spin channels: density ($d$), magnetic ($m$), singlet ($s$) and triplet ($t$) (For details of the spin-diagonalized notation used throughout this manuscript see e.g.~Ref.~\onlinecite{RohringerRMP}).
More complicated than the case of the Hubbard model, all three bare vertex functions become momentum-dependent.
As a result, the reducible vertex functions $\Phi_{d/m/s/t}$ depend not only on $q$ but also $k$ and $k^{\prime}$ even in the first iteration:
 \begin{align}
   \Phi^{kk'q}_{d/m} 
   &\to 
   \sum_{k_1} U^{\vek{k}\vek{k}_1\vek{q}}_{d/m} G_{k_1}G_{q+k_1}
      U^{\vek{k}_1\vek{k}'\vek{q}}_{d/m}
   \label{eq:1stredVerFctdm}\\
   \Phi^{kk'q}_{s/t}
   &\to \mp\frac{1}{2}
   \sum_{k_1} U^{\vek{k}\vek{k}_1\vek{q}}_{s/t} G_{k_1}G_{q-k_1}
      U^{\vek{k}_1\vek{k}'\vek{q}}_{s/t} \; .
   \label{eq:1stredVerFctst}
\end{align} 

The crucial observation of the {\it victory} implementation is the simpler structure of the reducible vertex functions which are used for extrapolation to reach the frequency-asymptotic of various two-particle vertex functions. 
In the Hubbard model, different momentum patches have essentially the same frequency-asymptotics~\cite{victory1} (known as kernel-I approximation in our notation) as can be seen from the only $q$ dependence of $\Phi_{d/m/s/t}$ calculated in the first iteration of Eqs.~(\ref{eq:1stredVerFctdm}) and (\ref{eq:1stredVerFctst}) if $U^{\vek{k}\vek{k}_1\vek{q}}$ is momentum independent for the Hubbard model, cf.~ Eqs.~(\ref{eq:PAd})-(\ref{eq:PAt}). 
With the presence of the $V$-interaction, this changes and $\Phi_{d/m/s/t}$ depend on all three momenta. 
As a result, the reducible vertex functions display a $\mathbf{k}, \mathbf{k^{\prime}}$-dependence even in the kernel-I function. 
We, therefore, modified the kernel-I function to faithfully incorporate such difference in different momentum patches. 
Instead of just taking one background value of the reducible vertex function, in each momentum patch (characterized by the different combinations of $\mathbf{k}$ and $\mathbf{k^{\prime}}$) we take a different background value to represent the kernel-I function of this momentum patch.  

The change of the bare vertex function also affects the evaluation of the self-energy $\Sigma_k$ in the Dyson-Schwinger equation
\begin{alignat}{1}
  \Sigma_k
     =& -\frac{1}{(N\beta)^2}\sum_{k'q}G_{k+q}G_{k'}G_{k'+q}
      \bigg[ 
      \frac{U}{2}\left[F_{d}-F_{m}\right]^{kk'q}\bigg. \nonumber\\
     &\bigg.+ V_{\vek{q}}\vesan{F}{kk'q}{d}
      \bigg]
     +\frac{1}{N\beta}\sum_{k'}G_{k}\left[U+2V_{\vek{q}=0}-V_{\vek{k}'-\vek{k}}\right]\;,
     \label{eq:SchwDys}
\end{alignat}
with $F$ being the full vertex in $d/m$ channels.
The other parquet equations formally remain unchanged.

Following Ref.~\onlinecite{PPPPaper}, we also improve the evaluation of Bethe-Salpeter equations by including, in the frequency part of the $k_1$ sum, the exactly known high frequency asymptotics of the full vertex $F$ and the irreducible vertex $\Gamma$~\cite{RohringerRMP,RohringerThesis}. 
A similar regularization (extension to high frequencies) is employed for the Dyson-Schwinger equation~\eqref{eq:SchwDys}. 
Furthermore, {\it victory} also benefits from the point group symmetry of the square lattice, where only the transfer momentum $\vek{q}$ in the irreducible Brillouin Zone (BZ) need to be stored in memory.
With these changes \victory can be used efficiently to study single-band correlated many-body systems with non-local interactions. 

\section{Results}
\label{sec:Results}

The evaluation of the parquet equations is numerically very demanding~\cite{victory1,victory2,Kauch2019} and, at the moment, only possible for small clusters. 
All the results presented in this work were obtained on a $6\times 6$ momentum cluster (corresponding to a $6\times 6$ cluster in real-space with periodic boundary conditions), which, nevertheless, represents the state of the art in the solution of the parquet equation.
Whenever possible, we used a finer grid for the Green's function, as explained in Ref.~\onlinecite{Kauch2019}. Since the bottleneck of the calculations is the memory consumption, only relatively small number of Matsubara frequencies can be taken, which restricts the computations to rather high  temperatures. Here the temperature was set to $T=t/6$ and $N_f=96$ Matsubara frequencies were used.   

\begin{figure*}[t]
\begin{minipage}{.58\linewidth}
   \centering
   \includegraphics[width=\linewidth]{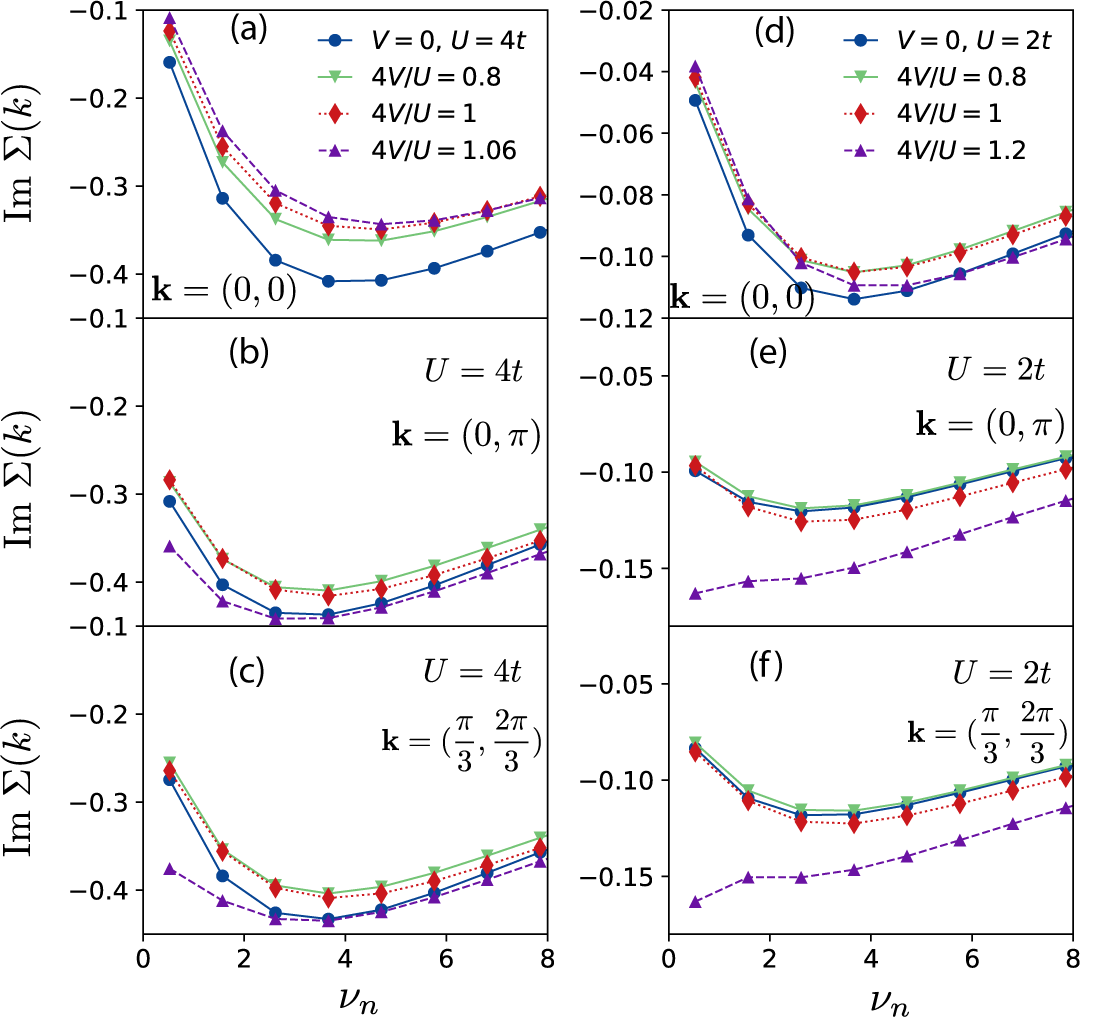}
   \caption{(a, d) Imaginary part of the self-energy $\Sigma_k$ as a function of the Matsubara frequency $\nu_n$ at $\vek{k}=(0,0)$  and the following  two $k$-points on the Fermi surface: (b, e) $(0,\pi)$ and (c, f) $\left(\frac{\pi}{3},\frac{2\pi}{3}\right)$.
   The first column displays the results for $U=4t$ and the second column for $U=2t$. Different symbols and colors denote different values of $V$. The largest value of $V$, indicated by a dashed line,  is close to the phase transition for a given $U$. }
   \label{fig:SigmaU24}
\end{minipage}%
\hfill
%
\begin{minipage}{.4\linewidth}
   \centering
   \includegraphics[width=\linewidth]{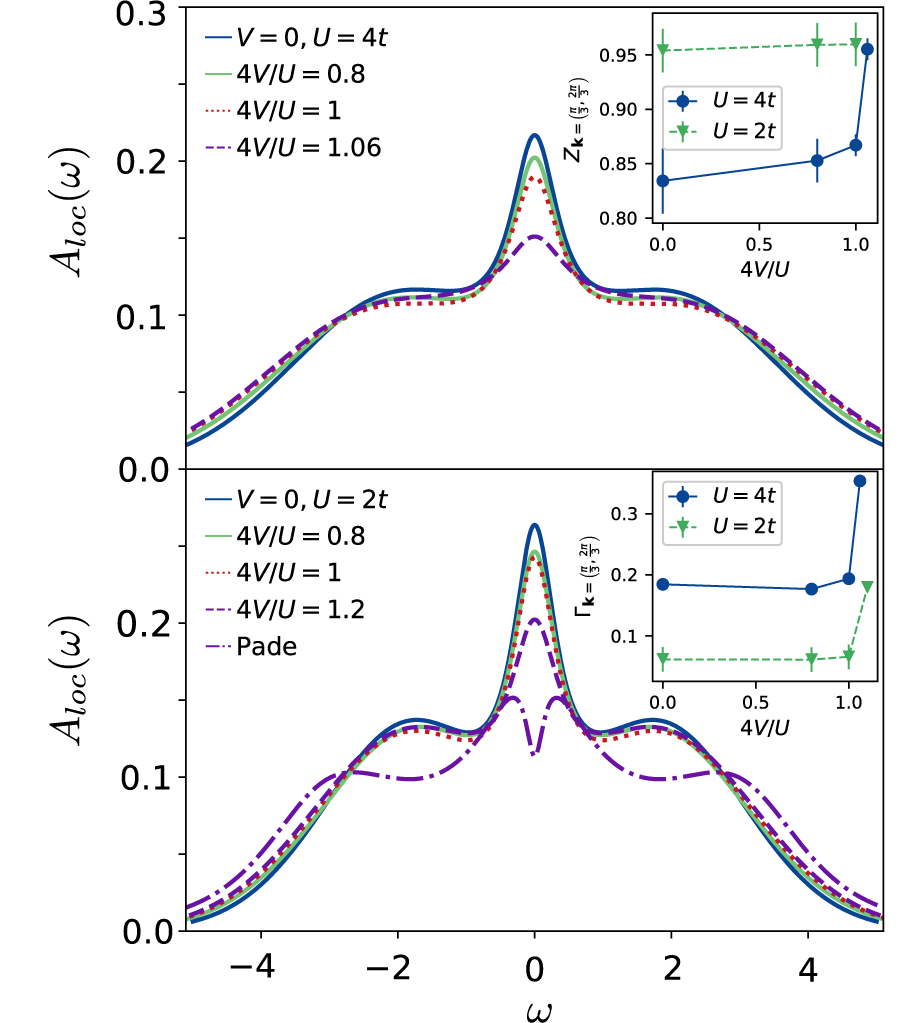}
   \caption{Local spectral functions $A_{loc}$ for different values of $V$ at $U=4t$ (top) and $U=2t$ (bottom) obtained by analytical continuation with MEM.
The violet dashed-dotted line in the bottom panel corresponds to a \pade{} interpolation.
The insets show Fermi liquid parameters as a function of $V$: the renormalization factor $Z_{\vek{k}}$ (upper inset) and  the scattering rate $\Gamma_{\vek{k}}$ (lower inset) for $\vek{k}=\left(\frac{\pi}{3},\frac{2\pi}{3}\right)$ which lies on the Fermi edge. The estimated error bars are also shown. }
   \label{fig:SpecFct}
\end{minipage}
\end{figure*}

\subsection{Single-particle self-energy and spectral function}
\label{sec:SelfEnergy}

In Fig.~\ref{fig:SigmaU24} we first show the single-particle self-energy $\Sigma(k)$ as a function of the Matsubara frequency. 
Two different values of $U$ were considered in these calculations, {\it i.e.} $U=4t$ in (a-c) and $U=2t$ in (d-f). 
Each row of Fig.~\ref{fig:SigmaU24} corresponds to a different $\vek{k}$-point in the BZ, {\it i.e.} $(0,0)$, $(0,\pi)$ and $\left(\frac{\pi}{3},\frac{2\pi}{3}\right)$. 
The latter two $\vek{k}$-points, {\it i.e.} $(0,\pi)$ and $\left(\frac{\pi}{3},\frac{2\pi}{3}\right)$, reside on the Fermi surface at half-filling. 
We vary the values of $V$ to probe the competition between the local and non-local Coulomb interactions in the EHM occurring around the transition boundary. 
We denote the results corresponding to different $V$ by different symbols and colors in Fig.~\ref{fig:SigmaU24}.

As can be clearly seen in Fig.~\ref{fig:SigmaU24}(a) and (d), away from the Fermi surface, the self-energy becomes smaller (in absolute terms) with the increase of $V$, in nice agreement with the expectation of the non-local interaction effect. 
As we know, if interaction between any two electrons becomes a constant regardless of their distance, the system is only determined by the single-particle hopping. The self-energy becomes frequency independent, {\em i.e.} a  constant for the occupied and unoccupied $\mathbf k$-points.
The EHM with only nearest-neighbor $V$ interpolates the Hubbard model and the above-mentioned infinite-range interaction model.
Consequently, it will display a finite but less correlated self-energy compared to that of the Hubbard model, cf.~Ref.~\onlinecite{PPPPaper}.

A similar reduction is observed for the points  on the Fermi surface, {\it i.e.} for $\vek{k}=(0,\pi)$ and $\vek{k}=\left(\frac{\pi}{3},\frac{2\pi}{3}\right)$, as long as  $4V/U<1$.
Calculations with parameters $4V/U=0.2$ and $4V/U=0.4$ at $U=4t$ and $U=2t$ (not shown here) also confirm this trend.

However, close to the transition boundary at $4V/U\sim1$ and $U=4t$ (and at a slightly smaller $V$ for $U=2t$) the self-energy changes dramatically, {\it i.e.} the first few lowest-frequency points of $\imag\Sigma$ clearly deviate from those for smaller $V$ values. They even start to increase in absolute values for $U=2t$ and $4V/U=1.2$, see Fig.~\ref{fig:SigmaU24}(e) and (f).
At  $4V/U=1.2$ and $U=2t$ the curvature of the self-energy changes completely, leading to a large negative $\imag\Sigma$ at $\nu_{n}\to0$. This indicates a pseudogap or at least  bad metallic paramagnetic phase close to the CDW transition.
As it will become clear later, this is due to the enhanced charge fluctuations caused by the non-local Coulomb interaction $V$ (cf. the eigenvalues and susceptibilities in \reff{EVSusQ}).
We hence interpret the large  negative $\imag\Sigma$ in Fig.~\ref{fig:SigmaU24}(e) and (f) as a precursor of the insulating CDW phase, which would be further enhanced at lower temperatures.  That is, we do not yet have CDW order yet but already long-range CDW correlations, which leads to a  (pseudo)gap similar as in case of antiferromagnetic fluctuations in the Hubbard model \cite{Schaefer_Fate2015}. 

Fig.~\ref{fig:SpecFct} shows the local spectral function $A_{loc}$ obtained from the $\vek{k}$-integrated spectral function 
$A_k=-\frac{1}{\pi}\imag G_{\vek{k}}(\omega)$ for real frequencies $\omega$.
The analytical continuation was done with the maximum entropy method (MEM)\cite{Jarrell1996133,JosefMEM}. 
In accordance with the self-energy calculations, $A_{loc}(\omega)$ is shown for the same parameters as in Fig.~\ref{fig:SigmaU24}.
For $4V/U=1.2t$ at $U=2t$ the \pade{} interpolation\cite{Vidberg} is also shown as it results in a slightly different spectrum.

Independent of the values of $V$,  at $U=4t$ $A_{loc}(\omega)$ consists of three contributions stemming from the dominant quasiparticle peak at zero frequency and the upper/lower Hubbard bands at $\omega\sim\pm U/2$ indicating the overall metallic nature of the solutions.
In case of $U=2t$, the peak positions of the Hubbard bands are slightly displaced at $\omega\sim\pm2t$.
Increasing $V$ suppresses and broadens the quasiparticle peak.
This suppression can be assigned to the  increase of charge fluctuations (cf. \reff{EVSusQ} below).
The effect  is however small for the case of $4V/U\lsim1$. 

For a better comparison, the Fermi liquid parameters, namely the renormalization factor $Z_{\mathbf{k}}=\left[ 1-\left.\imag\Sigma_k/\nu_n\right|_{\nu_n\to0} \right]^{-1}$ and the scattering rate $\Gamma_{\vek{k}}=\left|\imag\Sigma_k\right|_{\nu_n\to0}$, are extracted from the self-energy and are shown for  $\vek{k}=\left(\frac{\pi}{3},\frac{2\pi}{3}\right)$ in the upper and lower insets of \reff{SpecFct}, respectively. 
To do this,  the imaginary part of self-energy was fitted with a second order polynomial for the first three Matsubara frequencies and the error estimate was obtained from the comparison to a fit with a third order polynomial. The quasiparticle renormalization $Z_{\mathbf{k}=\left(\frac{\pi}{3},\frac{2\pi}{3}\right)}$
slightly grows with increasing $V$ up to $4V/U\sim1$ above which the growth becomes much sharper.
Similarly, the scattering rate $\Gamma_{\mathbf{k}=\left(\frac{\pi}{3},\frac{2\pi}{3}\right)}$ slightly decreases with increasing $V$ until $4V/U\sim1$. Afterwards, it substantially grows. 
The particular behavior of the Fermi liquid parameters is another evidence of the charge fluctuations that only appear in the close vicinity of the CDW phase transition.
For all other values of $V$ with $4V/U\lsim1$, the CDW fluctuations have very little impact on the one-particle properties. 

For $4V/U=1.2$ and $U=2t$, the spectral function is suppressed at small frequencies. For these parameters the analytic continuation by means of \pade{} interpolation resulted in a dip at the Fermi energy, cf. \reff{SpecFct}, indicating a tendency to opening of a gap. 
Despite the different solutions from the MEM and the \pade{} for this parameter, both qualitatively agree with the expectation of the charge fluctuations effect, and they are consistent with the large negative $\imag\Sigma$ at the Fermi surface (see \reff{SigmaU24}). 
We believe that with the further reduction of temperature a full gap can be obtained which smoothly evolves into the CDW gap in the ordered phase.

\subsection{Two-particle susceptibilities and eigenvalues}
\label{sec:EigenValAndSusc}

\begin{figure}[t]
   \centering
   \includegraphics[width=0.8\linewidth]{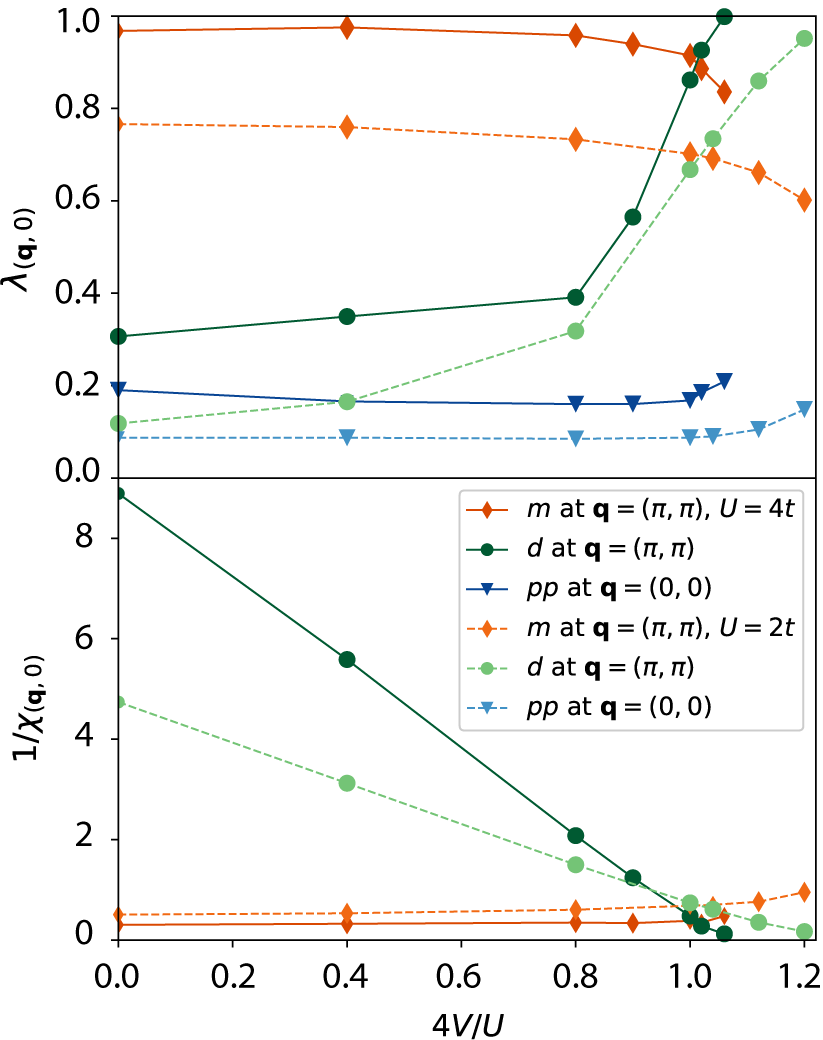}
   \caption{Leading eigenvalues (top) and inverse static susceptibilities (bottom) of the EHM within the PA for the density (green circles), magnetic (orange diamonds) and particle-particle (blue triangles) channel as a function of non-local interaction $V$.
   The dark symbols and solid lines correspond to data for $U=4t$ and the light symbols and dashed lines to $U=2t$, respectively.}
   \label{fig:EVSusQ}
\end{figure}

\begin{figure*}
   \centering
   \includegraphics[width=0.8\linewidth]{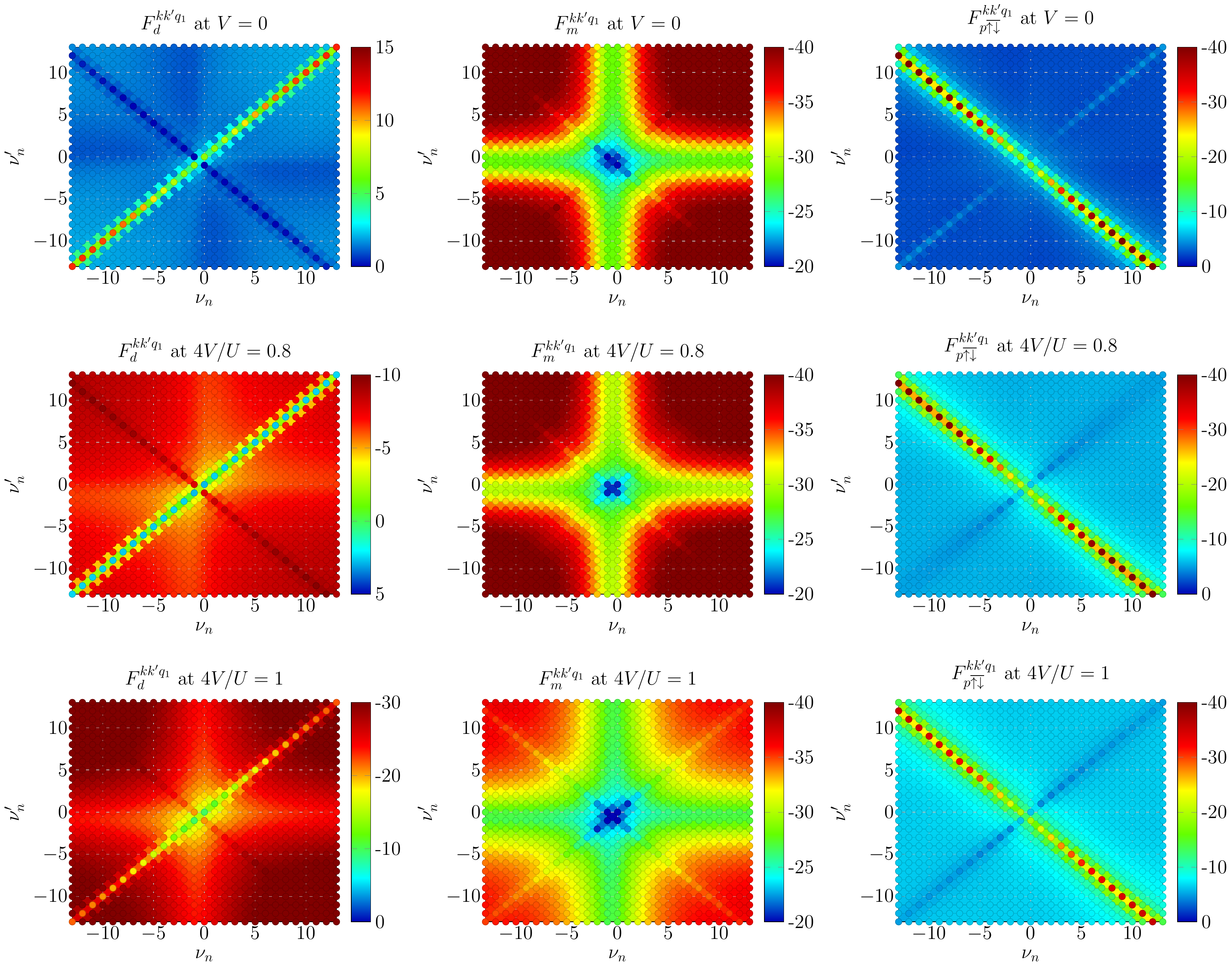}
   \caption{Full vertex $F^{kk'q_1}$ as a fuction of fermionic Matsubara frequencies $\nu_n$ and $\nu'_n$ for fixed momenta and bosonic frequency. Specifically, the four-vectors are set to $k=(\vek{k},\nu_n)=(0,0,\nu_n)$, $k'=(0,0,\nu'_n)$ and $q_1=(\pi,\pi,0)$. 
   Different columns correspond to the the density (first column), the magnetic (second) and the particle-particle (third) channels, respectively.
  Each row corresponds to the different values of the non-local interaction $V$:  $V=0$ (first row), $4V/U=0.8$ (second) and $4V/U=1$ (third);  the local interaction is set to $U=4t$.
   }
   \label{fig:Fnu}
\end{figure*}

A more direct measure of the fluctuations is the susceptibility in various channels.  
As the parquet approach determines the single- and two-particle quantities simultaneously, the signals of the CDW observed in the self-energy (Fig.~\ref{fig:SigmaU24}) and the spectral function (Fig.~\ref{fig:SpecFct}) will also leave their fingerprints in the two-particle vertex functions. 
Actually, as it will be seen soon the characterization of the phase transition can be more conveniently done in the two-particle level, as the two-particle susceptibilities are the fluctuations of the corresponding single-particle quantities.
These fluctuations become more significant in the vicinity of the phase transition, and are not directly accessible in any single-particle self-consistent theory. 
In the parquet approach, the susceptibilities in the  magnetic, density (charge) and particle-particle (including both singlet $s$ and triplet $t$) channels are given as
\begin{alignat}{1}
   \Chi_{d/m}(q) &=-\sum_k G_kG_{q+k} 
      \left[ 1 + \sum_{k'}F_{d/m}^{kk'q}G_{k'}G_{q+k'}\right]\label{eq:chi_dm}\;, \\
   \Chi_{pp}(q) &=-\sum_k G_kG_{q-k} 
      \left[ 1 + \sum_{k'}F_{p\overline{\up\dn}}^{kk'q}G_{k'}G_{q-k'}\right]\label{eq:chi_pp}\;,
\end{alignat}
with $F_{p\overline{\up\dn}}^{kk'q}=\frac{1}{2}[F_s - F_t]^{kk'q}$. 
The divergence of the susceptibility in a given channel signals the phase transition towards the breaking of the corresponding symmetry of this channel. 
Similarly, Eqs.~(\ref{eq:chi_dm}) and (\ref{eq:chi_pp}) also indicate that the phase transition can be equivalently characterized by the leading eigenvalue $\lambda$ of a corresponding Bethe-Salpeter equation which reaches the value of $1$. 
Therefore the eigenvalues themselves constitute another measure of the fluctuation strength in a given channel close to a phase transition.  The corresponding eigenvalue equations read 
\begin{alignat}{1}
   \lambda^{q}_{d/m}\phi_{d/m}^{kq} 
      &= \sum_{k_1}\vesan{\Gamma}{kk_1q}{d/m}
         G_{k_1}G_{k_1+q} \phi_{d/m}^{k_1q}\\
   \lambda^{q}_{pp}\phi_{pp}^{kq} 
      &= \sum_{k_1}
         \Gamma_{p\overline{\up\dn}}^{kk_1q}
         G_{k_1}G_{q-k_1}\phi_{pp}^{k_1q}\;,
   \label{eq:EVvsV}
\end{alignat}
where $\Gamma$ denotes the irreducible vertex in a given channel and $\phi$ is the eigenvector.
Since the Bethe-Salpeter equations are diagonal in transfer momentum $\vek{q}$ and frequency $\omega_n$,  the above eigenvalues $\lambda^{q}_{d/m/pp}$ also inherit this dependence. In the following we discuss eigenvalues only for the zeroth bosonic frequency ($\omega_n=0$), $\vek{q}=(\pi,\pi)$ for $\lambda_{d/m}$, and $\vek{q}=(0,0)$ for $\lambda_{pp}$, where the eigenvalues in the respective channels are dominant.

In \reff{EVSusQ} the eigenvalues (top) and (inverse) susceptibilities (bottom) are shown as functions of $V$.
Again results for two values of the local interaction are shown: $U=4t$ (solid lines) and $U=2t$ (dashed lines).  In the overview phase diagram \reff{PhaseWahle}, the ($U$, $V$) points of  \reff{EVSusQ} are marked as  upper and lower triangles.

Changes are visible in the charge susceptibility $\Chi_d$ which increases when $V$ is enhanced, but not surpassing the magnetic susceptibility $\Chi_m$.
The inverse of $\Chi_m$ is close to zero and likewise, its corresponding eigenvalue is  close to one, indicating very strong antiferromagnetic fluctuations ($\vek{q}=(\pi,\pi)$) in this parameter regime. For the smaller value of $U$ both AFM and CDW fluctuations are significantly weakened with  the AFM fluctuations being still dominant in the system.  

It is only at a rather large value of $V$, $4V/U> 0.8$ that the density eigenvalue  starts to rapidly increase (see the top panel of \reff{EVSusQ}). 
At the same time the magnetic eigenvalue slightly decreases with increasing $V$. 
When  $4V/U$ becomes slightly greater than $1$, the magnetic eigenvalue is eventually surpassed by the density eigenvalue. 
Only then the CDW fluctuations become more important, which can also be seen in the charge susceptibility that becomes bigger than the magnetic one when $4V/U\sim 1$ (see the bottom panel of \reff{EVSusQ}). 
For the smaller local interaction,  $U=2t$, the increase of the charge susceptibility and eigenvalue is less steep as compared to the case of $U=4t$ (cf. \reff{EVSusQ}). 
As a result, the dominance of CDW fluctuations occurs at a slightly larger $V$ value.
Nevertheless, for both cases ($U/t=2$ and $U/t=4$) the fluctuations in the charge channel become stronger than in the magnetic channel but only very close to the transition boundary. 
The leading eigenvalue and the near-divergence of the charge susceptibility occur at $\mathbf{q}=(\pi, \pi)$ which corresponds to a new periodicity in real-space of $\sqrt{2}\times\sqrt{2}$.  
This is in agreement with the strong coupling expansion for this model\cite{DongenStrCoup,Bari1971} and was also seen in forth-order perturbation theory.\cite{DongenStrCoup} 

The $pp$ eigenvalue and the corresponding susceptibility (not shown here) remain small for all values of $V$, which leads to a conclusion that, for the parameters considered, pairing fluctuations do not play any significant role. 
Note, however, that for $U=4t$ the $pp$ eigenvalue is almost twice as big as that for $U=2t$, {\it i.e.} a relatively strong local interaction is important for pairing fluctuations, as it is also for the AFM and CDW ones (cf. top panel of   \reff{EVSusQ}).

Although it is not the prime aim of this paper to precisely determine the transition boundary of the EHM, the above analysis on the leading eigenvalue and the susceptibility provide an efficient way to estimate the critical value of $V$. 
The value of $\lambda_d$ at $\vek{q}=(\pi,\pi)$ reaches almost $1$ for $4V/U\to1.06$ at $U=4t$ and for $4V/U\to1.2$ at $U=2t$, respectively.  
A more reliable estimate of the phase transition point within the parquet method is very difficult. It would require calculations for different and large cluster sizes to properly resolve the deviations from mean-field critical exponents.
Further, we are restricted to second order phase transitions because  the parquet equations are only solved in the paramagnetic phase. If there is a first order transition, as might well be the case in some parameter range \cite{Kapcia2017,Paki2019}, the thus calculated $V$ values are too large.

We note that very similar critical values of $4V/U$ for the CDW phase transition were obtained in many different works~\cite{Vanloon2014,Gull2017,Wahle1998,edmft1,Zhang1989}.
DQMC predicted a phase transition in the regime of $4V/U\in\{1,1.25\}$ for $U=4t$ and $4V/U\in\{1,1.4\}$ for $U=2t$~\cite{Zhang1989}.
In the DCA calculations~\cite{Gull2017}, they are $4V/U=1.04$ for $U=4t$ and $4V/U=1.216$ for $U=2t$ .
At a lower temperature ($T=t/12.5$), the dual boson approach~\cite{Vanloon2014} identifies the phase transition at $4V/U=1.08$ for $U=4t$ and $4V/U=1.04$ for $U=2t$.
Our estimation based on the parquet approximation is in agreement with these studies, confirming the feasibility of the parquet approach in the study of phase transitions. 

\begin{figure*}[htbp]
   \centering
      \includegraphics[width=0.8\linewidth]{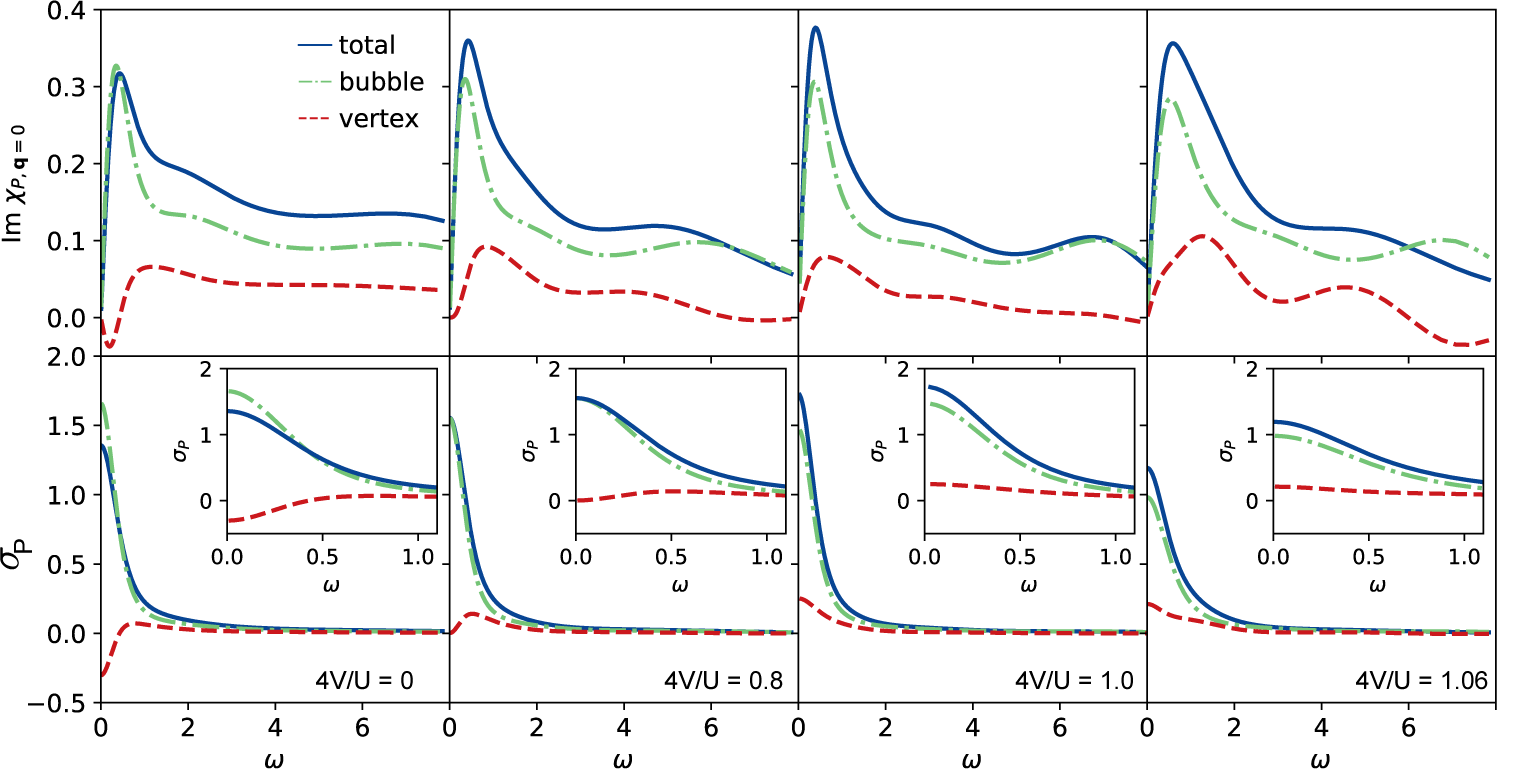} 
   \caption{Current-current correlation function (top row) and optical conductivity (bottom row) as responses to an external electric field $\vek{E}=E\,\vek{a}_x$ in the PA at four different non-local interactions $V$.
The response functions (dark blue solid line) are further decomposed into the bubble (light green dashed/dotted line) and the vertex (red dashed line) contributions.
The low-frequency zoom-in plots are shown in the insets for the optical conductivity.}
   \label{fig:optCondu_eCDW}
\end{figure*}

\subsection{Full \twop{} vertex function}
\label{sec:TPvertex}

In addition to the susceptibility and the leading eigenvalue, the competing fluctuations can also be seen in the full 
 \twop{} vertex function $F^{kk'q}$.
 In \reff{Fnu} we show the fermionic frequency dependence ({\it i.e.} as a function of $\nu_n$ and $\nu'_n$) of the full vertex in the density, magnetic and particle-particle channels (different columns) for different values of $V$ (different rows) at $U=4t$. The momenta are set to $\vek{k}=\vek{k'}=(0,0)$ (please note that the dependence on the fermionic momenta ${\bf k}$ and ${\bf k'}$ is very weak), $\vek{q}=(\pi,\pi)$ and the bosonic frequency $\omega_n=0$. 

Increasing the non-local interaction $V$ affects mostly the density channel. Comparing however the three rows of \reff{Fnu} we can conclude that the onset 
 of strong charge fluctuations is visible only shortly before the phase transition, {\it i.e.} at $4V/U\sim1$. For $V=0$, the full vertex function is mainly described by the main diagonal with $\nu_n=\nu'_n$. As $V$ is increased to $4V/U=0.8$, the overall amplitude of the full vertex function increases and becomes negative except for the main diagonal components.  
 This background, however, is nicely approximated by the first order contribution \refq{PAd},
$2V_{\vek{q}=(\pi,\pi)}-V_{\vek{k}'-\vek{k}=(0,0)}=-12V$, which for $4V/U=0.8$ is equal to $-9.6$ in our units.
When the charge fluctuations become stronger ($4V/U=1$, bottom row of \reff{Fnu}), these contributions significantly increase far beyond the first order approximation.
In the magnetic and the particle-particle channel (cf. the second and the third column of \reff{Fnu}) the structure of $F$ stays nearly unchanged compared to the density channel.
At $4V/U=1$ we see however a strong broadening of the plus-structure, {\it e.g.} along $\nu_0=\frac{\pi}{\beta}$ and $\nu'_0=\frac{\pi}{\beta}$ in the magnetic vertex $F_m$. This indicates a suppression of magnetic contributions due to the influence of the now very strong CDW fluctuations. 
In this regime, likewise, the magnetic eigenvalue and magnetic susceptibility decrease (see \reff{EVSusQ}).
Analyzing the momentum dependence of the full vertex function (not shown here) leads to the same conclusion. 
Predominant effects of the charge fluctuations are seen in $F_d$ only close to the phase transition for  $4V/U\gsim1$. If the phase transition was of first-order and hence at a smaller $V$, AFM fluctuations might even prevail throughout the entire paramagnetic phase up to the CDW phase transition line. 
Turning back to  the magnetic contributions: we observe a global suppression for $4V/U=1$, whereas for smaller non-local interactions this effect is very small.
The contributions of the particle-particle channel stay almost unchanged with the increase of the non-local interaction.

\section{Response to electric field}
\label{sec:OptCondu}
With the two-particle vertex function available in the parquet approach, we are also able to study the response to external perturbations easily.
Like in the two-particle susceptibility and the density-density vertex function, the non-local charge fluctuations are also encoded in  response functions which are experimentally accessible.
As an example, we apply an electric field along the $x$-direction to the EHM,  {\it i.e.} $\vek{E}=E\,\vek{a}_x$, and study the response  to it at different value of $V$. 
The response function (or paramagnetic current-current correlation function) within the Peierls approximation reads
\begin{alignat}{1}
   \Chi_{jj,q}
   =& \underbrace{-
    \frac{2}{(N\beta)^2}\sum_{kk'}
    \vesan{\gamma}{\vek{k}\frac{\vek{q}}{2}}{P}
    \vesan{\gamma}{\vek{k}'\frac{\vek{q}}{2}}{P}
      {G}_{k}{G}_{q+k}
      \vesan{F}{kk'q}{d}
      {G}_{q+k'}G_{k'}}_{
     \equiv\Chi_{jj,q}^{\ver}}\nonumber\\
   &\underbrace{-
    \frac{2}{N\beta}\sum_{k}
    \left[\vesan{\gamma}{\vek{k}\frac{\vek{q}}{2}}{P}\right]^2
      {G}_{k}{G}_{q+k}}_{
     \equiv\Chi_{jj,q}^{\bub}}
   \label{eq:CCP_optic}
\end{alignat}
for a coupling along the field direction, $x$, with
\begin{alignat}{1}
   \vesan{\gamma}{\vek{kq}}{P}
   &=2t\sin\left[\vek{a}_x(\vek{k}+\vek{q})\right]
   \equiv\frac{\partial}{\partial {\mathbf k}_x}\epsilon_{\vek{k}+\vek{q}}\;.
   \label{eq:gammaP}
\end{alignat}
Because of the long optical wavelength, the variation of electric field in space can be neglected.
Thus, it is sufficient to restrict the calculations to $\vek{q}=0$. The optical conductivity which is obtained generally from linear response theory as 
\begin{alignat}{1}
   \vek{j}_{\vek{q}=0}(\omega)
      &= \underbrace{
      \frac{\imag[\Chi_{jj,\vek{q}=0}(\omega+i\delta)
           -\Chi_{jj,\vek{q}=0}(i\delta)]
          }{\omega+i\delta}}_{\equiv\sigma(\omega)}\vek{E}_{\vek{q}=0}(\omega)\;.
   \label{eq:Condkw}
\end{alignat}
Following  Eq.~(\ref{eq:Condkw}), one can first transform the current-current correlation function from Matsubara to real frequency and then divide it with $
\omega+i\delta$ to get the optical conductivity.  
Alternatively, one can directly transform the $\imag\Chi_{jj,\vek{q}=0}(i\Omega_{m})/i\Omega_{m}$ to real frequency with MEM, which was the strategy taken in our calculations~\cite{JosefMEM}.  


The response fuctions to the applied electric field (in the direction of the lattice vector $\vek{a}_x$) are displayed on the real-frequency axis in \reff{optCondu_eCDW}.
The response functions $\Chi_{P,\vek{q}=0}$ and the optical conductivity $\sigma_P$, are displayed for various values of $V$ at $U/t=4$: from $4V/U\lsim1$ (weak $V$-limit) to $4V/U>1$ (strong $V$-limit) in which either antiferromagnetic or charge fluctuations prevail (cf. \refs{EigenValAndSusc}).
Additionally, the bubble and vertex contributions, defined in \refq{CCP_optic}, 
are displayed in Fig.~\ref{fig:optCondu_eCDW} by light green dashed/dotted lines and red dotted lines, respectively.

The correlation function exhibits a complex two peak structure with the maximum peak value increasing with $V$ up to $4V/U\sim1$ and decreasing afterwards.
However, there is only one peak observed in the optical conductivity.
The small high-frequency feature observed in $\mbox{Im}\chi_{P,q=0}$ is suppressed by the large frequency denominator leaving only the Drude-like peak visible in Fig.~\ref{fig:optCondu_eCDW}. 
We expect a more visible high-frequency  peak  when the temperature becomes lower or $U$ is increased so that pronounced and well separated Hubbard bands start to form.

We notice that the zero-frequency peak in the optical conductivity increases when increasing $V$ from $V=0$ up to $4V/U=1$.
Both the bubble and vertex contributions become larger.
In particular, the vertex contribution at $\omega=0$ changes its sign and continuously increases with the increase of $V$, indicating the correlation effect is always enhanced with increasing $V$. 
On the other hand, at $4V/U=1.06$ the zero-frequency peak in the optical conductivity clearly drops down, which mainly results from the decrease of the bubble contributions. 
The vertex part, however, still strongly contributes.
The  bubble susceptibility in turn contributes less and less due to the suppression of the single-particle density of states at the Fermi level, see Fig.~\ref{fig:SpecFct}. 
As already discussed in the context of the self-energy and one-particle spectrum, this is because of the emergence of long-range CDW fluctuations at $4V/U > 1$ as is also reflected in the dramatic increase of the CDW susceptibility in Fig.~\ref{fig:EVSusQ}.
For the optical conductivity calculations, we observe the same effect as discovered in the self-energy and two-particle vertex functions: only in the close vicinity of the phase transition (around $4V/U\sim1.06$ at $U=4t$), the effect of the charge fluctuations becomes dominant. 
Away from the phase transition, for this half-filled two-dimensional square lattice, spin fluctuations are always dominant.

\section{Conclusion}
\label{sec:Conclusion}
By extending the parquet approach to include the non-local Coulomb interactions, we studied the extended Hubbard model on a two-dimensional square lattice using the parquet approximation.   
Through the calculations of the single-particle self-energy (Fig.~\ref{fig:SigmaU24}), spectral function (Fig.~\ref{fig:SpecFct}) and the two-particle susceptibility (Fig.~\ref{fig:EVSusQ}), the vertex function (Fig.~\ref{fig:Fnu}),  as well as the optical conductivity (Fig.~\ref{fig:optCondu_eCDW}), we are able to consistently identify the strong competition of the non-local charge fluctuations and the spin fluctuations. Quite surprisingly, the spin fluctuations are dominating over a wide range of parameters and only in the immediate  vicinity of the phase transition to the CDW order charge fluctuations prevail. In case of a first order transition \cite{Kapcia2017,Paki2019}, AFM  fluctuations might even dominate in the entire paramagnetic phase up to the CDW phase.

In the narrow regime, where the charge susceptibility surpasses the antiferromagnetic susceptibility, the self-energy in the paramagnetic phase turns from a metallic to an insulating-like behavior --- the vanguard of the CDW insulator.  Likewise the optical conductivity is suppressed by the strong CDW fluctuations. The parquet approach alows us to look at this competition and suppression also on a deeper level, unseen directly in the measured quantities. The otherwise very sharp frequency structure of the two-particle magnetic vertex becomes broadened close to the transition and the magnetic fluctuations are suppressed by the very large vertex in the charge channel. For the pairing (particle-particle) fluctuations, on the other hand, this has very little influence.

Due to the fact that the self-consistency at both, one- and two-particle, levels is simultaneously satisfied in the parquet approach, it is not surprising to obtain a consistent picture for all the above-mentioned quantities. 
Although it is not the prime interest of this work to locate the precise transition boundary, our rough estimation  agrees with the published results.  
Our work provides a valuable explanation of how the single- and two-particle quantities are internally related. 
The transparent structure of the parquet equation provides a predicting power in the sense that the change in one quantity can be naturally evolved to interpret the change of other quantities.

\section{acknowledgment}
We would like to thank Josef Kaufmann for the help with the analytical continuation.
P.P. has been supported by the 
Austrian Science Fund (FWF) through the 
Doctoral School ``Building Solids for Function'', A.K. and K.H. through FWF   project P 30997 and P 32044. G.L. acknowledges the starting grant of ShanghaiTech University, the Program for Professor of Special Appointment (Shanghai Eastern Scholar) and the support from the National Natural Science Foundation of China (Grant No. 11874263).
Calculations have been done in part on the Vienna Scientific Cluster (VSC).

\bibliographystyle{apsrev4-1}
%

\end{document}